\documentclass[a4paper]{aa}
\usepackage{natbib}
\usepackage{epsfig}
\bibpunct{(}{)}{;}{a}{}{,} 

\begin{document} 

\newcommand{\es}{erg s$^{-1}$}   
\newcommand{\ecms}{erg~cm$^{-2}$~s$^{-1}$}
\newcommand{\halpha}{H$\alpha$}  
\newcommand{\hbeta}{H$\beta$}
\newcommand{\kms}{km~s$^{-1}$}   
\newcommand{\cmthree}{cm$^{-3}$}
\newcommand{\msun}{M$_{\odot}$} 
\newcommand{\xmm}{XMM-\emph{Newton}} 
\newcommand{\nh}{\mbox{$N({\rm H})$}}
\newcommand{\chandra}{\emph{Chandra}}

\title{A large X-ray flare from the Herbig Ae star V892~Tau}
\author{G.\ Giardino\inst{1} \and F.\ Favata\inst{1} \and G.\ 
  Micela\inst{2} \and F. Reale\inst{3}} 

\institute{Astrophysics Division -- Research and Science Support
  Department of ESA, ESTEC, 
  Postbus 299, NL-2200 AG Noordwijk, The Netherlands
\and
  INAF -- Osservatorio Astronomico di Palermo, 
  Piazza del Parlamento 1, I-90134 Palermo, Italy 
\and
  Universit\`a di Palermo -- Dipartimento di Scienze Fisiche ed
  Astronomiche, Piazza del Parlamento 1, I-90134 Palermo, Italy 
}

\offprints{G. Giardino,\\ ggiardin@rssd.esa.int}

\date{Received date / Accepted date}

\titlerunning{}
\authorrunning{G. Giardino et~al.}

\abstract{We report the \xmm\ observation of a large X-ray flare from
  the Herbig Ae star V892~Tau. The apparent low mass companion of
  V892~Tau, V892~Tau~NE, is unresolved by \xmm. Nevertheless there is
  compelling evidence from combined \xmm\ and \chandra\ data that the
  origin of the flare is the Herbig Ae star V892~Tau.  During the
  flare the X-ray luminosity of V892~Tau increases by a factor of
  $\sim 15$, while the temperature of the plasma increases from $kT
  \simeq 1.5$~keV to $kT \simeq 8$~keV. From the scaling of the flare
  event, based on hydrodynamic modeling, we conclude that a 500 G
  magnetic field is needed in order to confine the plasma.  Under the
  assumptions that a dynamo mechanism is required to generate such a
  confining magnetic field and that surface convection is a necessary
  ingredient for a dynamo, our findings provide indirect evidence for
  the existence of a significant convection zone in the stellar
  envelope of Herbig Ae stars.  }

\maketitle

\section{Introduction}
\label{sec:intro}

Understanding the genesis and early evolution of intermediate mass
stars is a fundamental problem in studies of star formation. Owing to the
differences in stellar and circumstellar physics, as well as in
time scales, the evolution of intermediate pre-main-sequence stars is
qualitatively different from that of lower- and higher-mass stars.
The past 10 years or so have witnessed an increased interest in the
subject, and a key issue is the nature and evolution of Herbig Ae/Be
stars. These pre-main-sequence stars, with masses ranging between
about 2 and 10 $M_{\sun}$, are the more massive counterparts of T
Tauri stars.  They share with lower mass T Tauri stars peculiarities
such as infrared excess emission, conspicuous (optical and UV) line
emission and irregular photometric variability (see \citealp{ww98} for
a review).

According to classical models (\citealp{iben65}), pre-main-sequence
stars with masses in excess of 2 ${\rm M}_{\sun}$ are expected to
follow fully radiative tracks once the quasi-static contraction has
ended. The detection of Herbig Ae/Be stars in X-rays (\citealp{zp94};
\citealp{dms+94}) was therefore somewhat unexpected. Main sequence
stars of the same class (from later B to later A stars) do not possess
either a strong stellar wind nor a corona (as no magnetic dynamo
mechanism is available).  In contrast Herbig Ae/Be stars appear to
possess a strong stellar wind (\citealp{sbs93}; \citealp{bcs97}) and a
magnetic dynamo associated with an outer convection zone has been
invoked to explain the periodic variation of the emission lines
(\citealp{pcs+86}) as well as the stellar wind itself
(\citealp{fm84}).

\cite{zp94} proposed that the observed X-ray luminosity is linked to
the stars' strong stellar winds, possibly originating in shocks due to
wind instabilities and/or in the collision between the fast wind and
the remnant circumstellar material. \cite{dms+94} also favored a
stellar wind origin for the X-ray emission from Herbig Be stars.
Nevertheless a magnetically heated corona could also be at the origin
of the observed X-ray emission (\citealp{zp94}).


The presence of winds and the possibility of magnetic activity make
these objects interesting targets for X-ray studies. The unambiguous
detection of flaring activity in any Herbig Ae/Be star would be
significant as it would provide indirect but strong evidence for the
presence of a magnetically confined corona and thus of an operational
dynamo mechanism.

So far there is only one reported observation of flaring activity in
an Herbig Be star: \cite{htb+00} performed ASCA observations of the
Herbig Be star MCW~297, and reported the detection of a large flare
during which the X-ray luminosity of the source increased by a factor
of 5, with the plasma temperature increasing from 2.7~keV during the
quiescent phase to 6.7 keV at flare maximum. The interpretation
however is affected by some ambiguity due to the large ASCA point
spread function (PSF). \cite{tpn98} found about 20 infrared sources in
the ASCA error circle around MWC~297, the majority of which are likely
to be low-mass protostars. The peak X-ray luminosity of $\simeq
5\times 10^{32}$~\es\ reported by \cite{htb+00} would correspond to a
very large -- but still possible -- flare for a typical low-mass
active T Tauri.  For example \cite{tkm+98} have reported ASCA
observations of a large flare from the Weak-lined T Tauri star (WTTS)
V773~Tau, in which the peak flare luminosity was at least $\sim
10^{33}$~\es. The MWC~297 event therefore would not be exceptional for
a low mass pre-main-sequence star. \cite{hyk02} have recently
re-observed MWC~297 with \chandra, finding the source hundred times
less luminous in X-rays than estimated from ASCA observation, and with
no evidence for variability, thus casting some doubts on the original
interpretation.

In this article we report the observation of a large X-ray flare from
the Herbig Ae star V892~Tau.  We have used publicly available \xmm\
and \chandra\ data to monitor the X-ray activity of V892~Tau and
during one of the two \xmm\ exposures a large flare is observed. While
the apparent companion of V892~Tau, V892~Tau~NE, is unresolved by the
\xmm\ PSF, as discussed later, the evidence that the observed flare is
coming from the Herbig Ae star is compelling. 

The present paper is organized as follow: after a brief introduction
of the properties of V892~Tau below, the observations are described in
Sect.~\ref{sec:obs}. The spectral and timing analysis of the data are
presented in Sect.~\ref{sec:analysis} and the results are discussed in
Sec.~\ref{sec:disc}.

\subsection{The Herbig Ae star V892~Tau}

V892~Tau -- also known as Elias 1 -- is a young stellar object located
in the Taurus dark cloud complex, and is supposed to be the source of
the illumination for the faint nebula IC 359. The apparent magnitude
of V892~Tau is $R = 13.14$ (\citealp{ss94}) and its spectral
classification varies from B9 (\citealp{ss94}) and A0
(\citealp{elias78}; \citealp{fm84}) to A6 (\citealp{ck79},
\citealp{bci+92}, \citealp{twp94}).  Estimates for the visual
extinction also vary from $A_V \sim 8$ (\citealp{elias78}) and $A_V =
8.85$ (\citealp{ss94}, derived from a simultaneous estimate of the
spectral type and of the apparent color, $R-I = 1.71$) to $A_V \sim
3.9$ (\citealp{zp94}). The star is usually placed at a distance of
140~pc because of its association with the Taurus dark clouds
(\citealp{elias78}). The bolometric luminosity of V892~Tau is $L\sim
38~L_{\sun}$ (\citealp{bci+92}) and the source is variable in the
near-infrared (\citealp{elias78}).

Through near-infrared speckle interferometry \cite{km91} resolved
V892~Tau into an unresolved core and a sub-arcsec structure elongated
in the east-west direction. They interpret the light of the elongated
structure as being reflected within an edge-on circumstellar disk of
moderate optical depth. Newer near-infrared speckle interferometry
observations were performed by \cite{hlr97}, who favor a scenario in
which the diffuse component is due to scattering in bipolar lobes with
a polar axis oriented east-west.

V892~Tau appears to be a binary system (\citealp{lrh97}). The apparent
stellar companion, hereafter referred to as V892~Tau~NE, lies $4.1$
arcsec to the northeast, at position angle $22^{\circ}$\footnote{In
  the convention adopted by Skinner et al.~(1993), position angle
  $0^{\circ}$ is north and position angle $90^{\circ}$ is east.}
(\citealp{sbs93}). The available measurements allow a tentative
classification of V892~Tau~NE as a WTTS with spectral type M2 reddened
by 8--12 mag of visual extinction (\citealp{lrh97}). The study by
\cite{lrh97}, based on speckle interferometry, was carried out with
the explicit purpose of detecting binaries among Herbig Ae/Be stars.
\cite{lrh97} achieved a resolution of $\sim 0.1$ arcsec, that at a
distance of 140 pc correspond to 14 AU, and they do not detect other
stars in the vicinity of V892~Tau.  They conclude that in the
near-infrared V892~Tau is basically a wide binary system.

\section{Observations}
\label{sec:obs}

The X-ray observations discussed in this paper were obtained with
the \xmm\ and the \chandra\ observatories.  The \xmm\ observations of
V892~Tau consists of two deep (74.4 and 45.1 ks nominal) consecutive
exposures, the first starting on March 11 2001 at 12:40:22 UT and the
second one starting on March 12 2001 at 10:23:10 UT. All three EPIC
cameras were active at the time of the observations, in full-frame
mode with the medium filters.  The Principal Investigator for these
observations is F. Walter and the observation target is the triple
WTTS system V410 Tau. The observations are publicly available from the
\xmm\ archive.

The raw \xmm\ data have been processed by us with the standard SAS
V5.4.1 pipeline system, concentrating, for the spectral and timing
analysis, on the EPIC-pn camera. In each of the two \xmm\ exposures
the background is affected by a large proton flares of more than 10 ks
of duration. We have retained only time intervals in which the count
rate for the whole frame of photons above 5 keV was below a certain
threshold (3.3 cts/s in the present case). This operation omits
roughly 30\% of the observing time, but effectively reduces the
background level by a factor of $\simeq 4$. Source and background
photons were extracted using a set of scripts purposely developed at
Palermo Observatory.

Source and background regions were defined interactively in the
\textsc{ds9} display software, with the background extracted from
regions on the same CCD chip and at the same off-axis angle as for the
source region.  Response matrices (``\textsc{rmf} and \textsc{arf}
files'') appropriate for the position and size of the source
extraction regions were computed.  The spectral analysis has been
performed using the XSPEC package, after rebinning the source spectra
to a minimum of 20 source counts per (variable width) spectral bin.

The \emph{Chandra} ACIS observation of V892~Tau was taken starting on
March 7 2002 at 6:15:28 UT (18.0 ks nominal). The Principal
Investigator for these observations is P. Predehl and also in this
case the observation target is V410 Tau\footnote{There are other two
  \chandra\ exposures of V410 Tau available in the public archive (of
  15 and 11 ks respectively). In these observations however V892 Tau
  is at an extreme off-axis angle, and we have therefore not analysed
  them.}. The data were retrieved from the public data archive, with
no re-processing done on the archival data. Source and background
regions were defined in \textsc{ds9}, and light curves and spectra
were extracted from the cleaned photon list using CIAO V.~2.2.1
threads, which were also used for the generation of the relative
response matrices. Spectral analysis was performed in \textsc{xspec}
in the same way as for the \xmm\ spectra.

\begin{figure*}[!tbp]
  \begin{center} \leavevmode 
	\epsfig{file=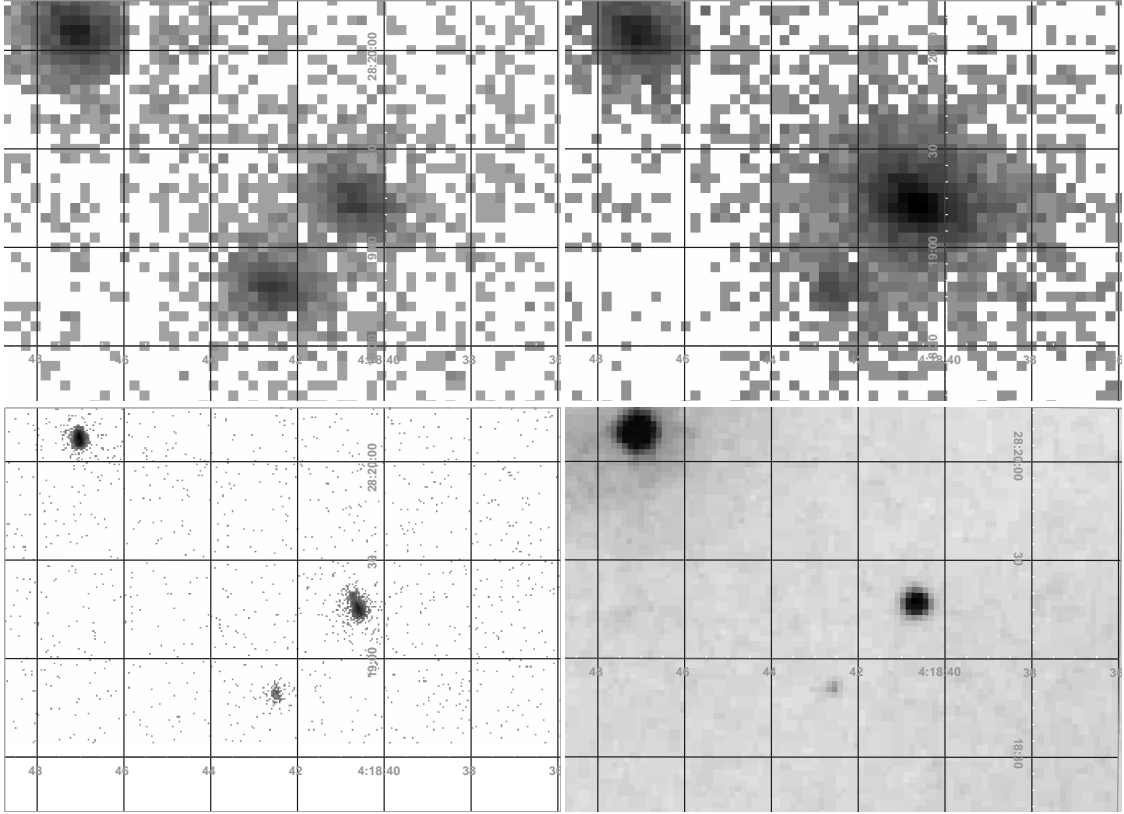, width=15.0cm}

\caption{The field centered on the system V892~Tau and V892~Tau~NE in
  the EPIC-pn camera before the flare (top left) and during the flare
  (top right), in the ACIS camera (bottom left) and in the Digital Sky
  Survey (bottom right). The images are on the same coordinate scale.
  The system is resolved in the \chandra\ data. The source at the
  south-east of the V892~Tau system is [BH98] MHO~11, a T Tauri star,
  the source at North-East is Hubble~4, also a T Tauri.}

  \label{fig:image} \end{center}
\end{figure*}

Fig.~\ref{fig:image} shows the images of the V892~Tau system in the
\chandra\ ACIS camera and \xmm\ EPIC-pn camera (before and after the
large flare) as well as the Palomar Digital Sky Survey image of the
field.  The 4 images are on the same sky coordinate scale.  The source
at the centre of the images is the V892~Tau system.  V892~Tau and
V892~Tau~NE are clearly resolved in the \chandra\ observations. The
\xmm\ point spread function has a full width at half maximum of $15$
arcsec and therefore cannot resolve the system. The source at $38$
arcsec to the SE of the V892~Tau system is [BHS98] MHO 11, a T Tauri
star (\citealp{bhs+98}), first identified in ROSAT data by
\cite{ss94}. The bright source in the North-East corner is Hubble~4, a
well known T Tauri.

In Table~\ref{tab:coord} we report the coordinates of V892~Tau and
V892~Tau~NE as derived from the radio (VLA) observation of
\cite{sbs93}, to be compared with the coordinates of the sources in
the \xmm\ and \chandra\ data.  The source coordinates from the EPIC-pn
data correspond to the peak of a gaussian distribution fitted to the
source image, the source coordinates for the \chandra\ data simply
correspond to the brightest pixel in the source image. The agreement
between the sources' positions in the \chandra\ image and as
determined from VLA observations is excellent (within 0.4 arcsec). The
source coordinates derived from the \xmm\ observations before the
flare have a 2.4 arcsec offset from V892~Tau (well within the
uncertainty expected for the determination of positions of EPIC X-ray
sources) and a 6 arcsec offset from V892~Tau~NE. After the flare the
position offsets become 3.1 arcsec and 6.8 arcsec from V892~Tau and
V892~Tau~NE respectively.




\begin{table*}[thbp]
  
   \caption{Positions for the two components of the V892~Tau
     system from radio (VLA) observations (Skinner~et~al.~1993) and
     as derived here from \xmm\ and \chandra\ images.} 
\begin{center} 
\leavevmode
    \begin{tabular}{lcccc}
Source 	& VLA	&  XMM$^a$ & XMM$^b$ & Chandra\\
\hline
~	& {\tiny  RA(J2000)} & {\tiny  RA(J2000)} & {\tiny  RA(J2000)} & {\tiny  RA(J2000)}\\
~	& {\tiny Dec(J2000)} & {\tiny Dec(J2000)} & {\tiny Dec(J2000)} & {\tiny Dec(J2000)}\\
\hline
V892~Tau & 4 18 40.60 &  4 18 40.67  & 4 18 40.67 & 4 18 40.63\\
~	 & 28 19 15.9 & 28 19 13.7  & 28 19 12.9 & 28 19 15.8\\
V892~Tau~NE & 4 18 40.70 &  $^c$ & $^c$ & 4 18 40.74\\
~	    & 28 19 19.7 & $^c$ & $^c$ & 28 19 19.3\\
    \end{tabular}
    \label{tab:coord}
  \end{center}
	$^a$X-ray quiescent, $^b$X-ray flaring, $^c$unresolved from
        V892~Tau. 
\end{table*}



\section{Spectral and timing analysis of V892~Tau and V892~Tau NE}
\label{sec:analysis}

\subsection{\emph{Chandra} observation}

\begin{figure}[!tbp]
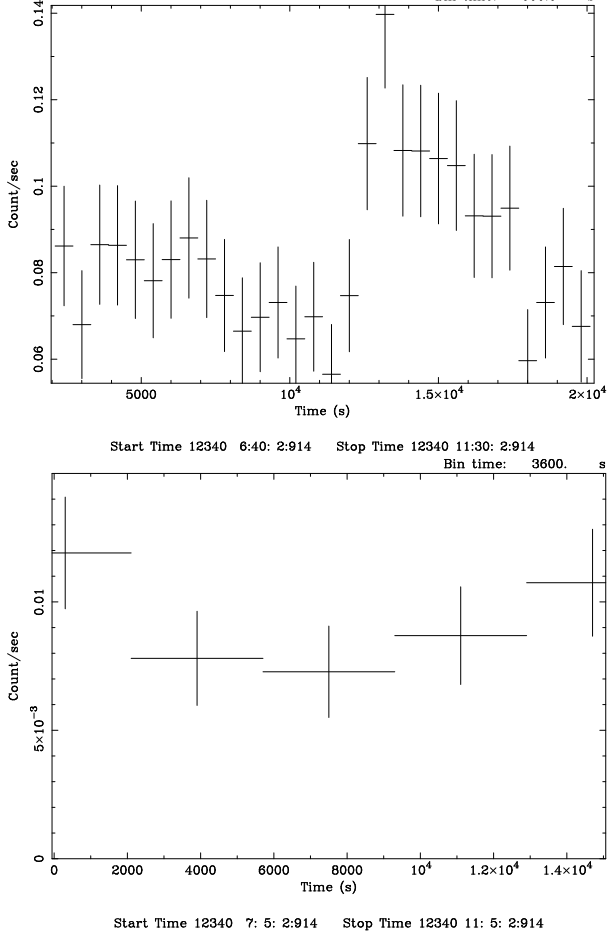

  \begin{center} \leavevmode 
	\epsfig{file=151f2.ps, height=8.0cm, angle=270}
	\epsfig{file=151f3.ps, height=8.0cm, angle=270}

\caption{Background-corrected light curves of V892~Tau (top) and
  V892~Tau~NE (bottom) from the \chandra\ data, where the two stars
  are well resolved. Note the different vertical scale.}

  \label{fig:lc_chandra}
  \end{center}
\end{figure}

The light curves of V892~Tau and V892~Tau~NE during the $\sim$ 18~ks
\chandra\ observation of March 2003 are shown in
Fig.~\ref{fig:lc_chandra}. In the \chandra\ data, The Herbig Ae star
V892~Tau is clearly resolved from the less luminous companion, the
WTTS V892~Tau~NE.

During the \chandra\ observation V892~Tau shows significant
variability, with its X-ray luminosity varying by a factor of 2 in
less than 1 ks. The rise in the source luminosity is impulsive (the
source doubles its luminosity in less than 1 ks) and it is followed by
a slow decay.  A proper time resolved spectral analysis of the rise
phase is not possible with the signal to noise ratio of this data set.
We have inspected the light curves of the source in a soft band
($0.0$--$1.7$~keV) and in a hard band($1.7$--$8$~keV) and see evidence
for a hardening of the spectrum during the rise phase. The decay time
$\tau \simeq 1$~hr is typical of X-ray stellar flaring events.

The bottom panel of Fig.~\ref{fig:lc_chandra} shows the light curve
of V892~Tau~NE. The source is weak, on average 8 to 10 times less
luminous than V892~Tau. Within the uncertainty of the large
error--bars the source does not appear significantly variable. The
probability of constancy of V892~Tau~NE according to the
Kolmogorov-Smirnov test (which measures the maximum deviation of the
integral photon arrival times from a constant source model) is 21\%.

\begin{figure}[!tbp]
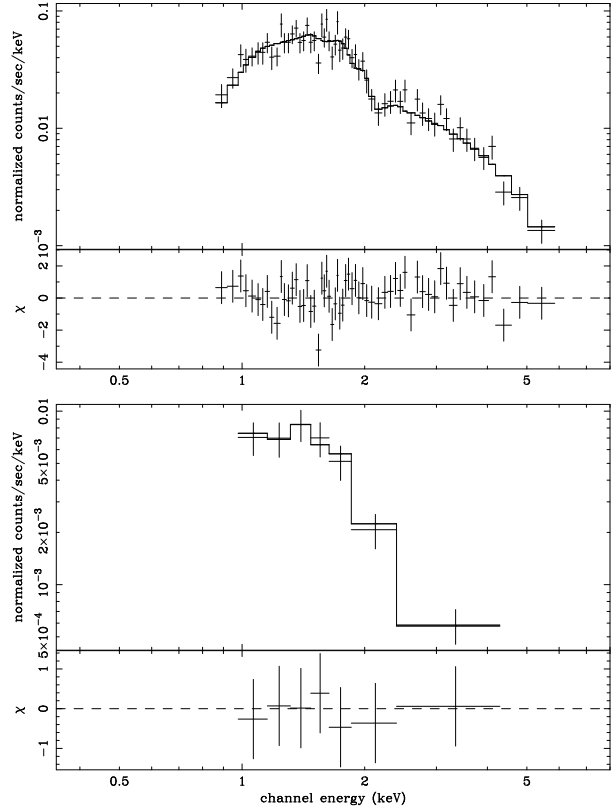

  \begin{center} \leavevmode 
	\epsfig{file=151f4.ps, height=8.0cm, angle=270, bbllx=100, bblly=43, bburx=535, bbury=700, clip=}
	\epsfig{file=151f5.ps, height=8.0cm, angle=270, bbllx=100, bblly=43, bburx=554, bbury=700, clip=}

\caption{Background-corrected spectra of V892~Tau (top) and
  V892~Tau~NE (bottom), from the \chandra\ data. The best-fit absorbed
  1-$T$ models are also shown.}

  \label{fig:ps_chandra}
  \end{center}
\end{figure}

The ACIS spectra of V892~Tau and V892~Tau~NE are shown in
Fig.~\ref{fig:ps_chandra}. As summarised in Table \ref{tab:spectra}
the spectrum of V892~Tau is well fitted by an absorbed 1-T plasma
model with $N(H) = 0.83\pm0.08 \times 10^{22}~{\rm cm^{-2}}$ and
temperature $kT = 2.10\pm0.19$~keV. The metallicity of the plasma is
not well constrained by the data. The spectral data for V892~Tau~NE
have poor signal to noise ratio, nevertheless a fit with an absorbed
1-T plasma model provide a useful estimate of the absorbing column
density and the plasma temperature. We derive a value of $N(H) =
1.20\pm0.18 \times 10^{22}~{\rm cm^{-2}}$, which is similar to the
value derived for V892~Tau and therefore consistent with the
hypothesis that V892~Tau~NE is a physical companion of V892~Tau. The
derived best-fit value of $kT = 1.04\pm0.17$~keV for the plasma
temperature of V892~Tau~NE is instead significantly different from the
value derived for V892~Tau.  The model dependent fluxes (as derived
from the best-fit models) are $7.1 \times 10^{-13}$\ecms\, in the band
$0.55$--$7.50$~keV, and $5.7\times 10^{-14}$\ecms, in the band
$0.67$--$7.50$~keV, respectively for V892~Tau and V892~Tau~NE.

\subsection{\xmm\ observations}

\begin{figure}[!tbp]
  \begin{center} \leavevmode 

	\epsfig{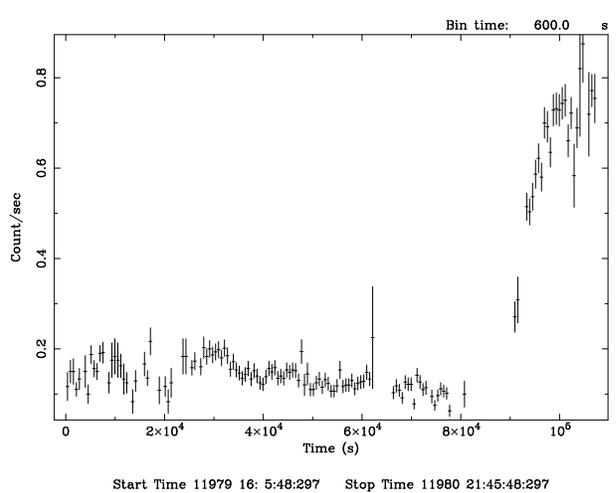}

\caption{Background-subtracted light curve of the system V892~Tau,
  from the two consecutive \xmm\ observations. With \xmm\ the Herbig
  Ae star V892~Tau is unresolved from the companion V892~Tau~NE,
  nevertheless our data indicate that this light curve is completely
  dominated by the activity of V892~Tau (see discussion).}

  \label{fig:lc_xmm}
  \end{center}
\end{figure}

The light curve of the V892~Tau system derived from the two
consecutive \xmm\ exposures are shown in Fig.~\ref{fig:lc_xmm}. All
the gaps in the light curve except the one around 64 ks are due to
the filtering process that we have applied to the raw data in order to
remove the effect of solar proton flares. In particular roughly 10 ks
at the beginning of the first exposure have been removed. The gap
around 64 ks is due to the time difference between the end of the
first exposure and the start of the second (one hour).

During the first $\sim 80$ ks the light curve of the
V892~Tau-V892~Tau~NE system presents significant variability with a
time scale and amplitude similar to the one observed for V892~Tau in
the \chandra\ data. During the last 30 ks the light curve of the
system undergoes a dramatic variation, in what appears to be a large
flare. The source counts increases by a factor of $\sim 8$ in less
than 10 ks. The source flux then remains at around the maximum level
until the end of the \xmm\ exposure.

We have performed the spectral analysis of the two \xmm\ exposures
separately and we summarise the results 
into two separate tables: Table~\ref{tab:xmm74ks_ps} for the first
exposure and Table~\ref{tab:spectra} for the second exposure. The
reason for this is that while for the first exposure two temperature
plasma models are necessary to obtain acceptable fits to the spectra
for the second exposure one-temperature plasma models provide adequate
fits to the spectral data.

\begin{figure}[!tbp]
  \begin{center} \leavevmode 
	\epsfig{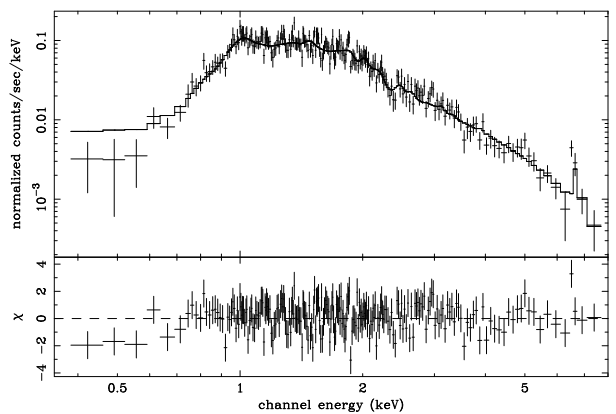}

\caption{Background corrected spectrum of the V892~Tau system during
  the first \xmm\ observation (before the flare).}

  \label{fig:ps_xmm74ks}
  \end{center}
\end{figure}

The spectrum of the V892 Tau system during the first exposure of 74
ks, corresponding to an effective integration time of $\la 64$~ks
(after the proton flare filtering process), is shown in
Fig.~\ref{fig:ps_xmm74ks}, together with the best fit absorbed 2
temperature plasma model. The best fit values for the model parameters
are summarised in Table~\ref{tab:xmm74ks_ps} together with the reduced
$\chi^2$ of the fit and its null hypothesis probability, $P$. A model
with an absorbing column density of $N({\rm H}) = 0.92 \times
10^{22}$~cm$^{-2}$, plasma temperatures $kT_1 = 1.02$~keV and $kT_2=
2.82$~keV and a metal abundance $Z = 0.21$ provides a good fit to the
integrated spectrum. An absorbed one temperature plasma model (as used
for the \chandra\ data, which however have lower statistics) does not
provide an acceptable description of the spectrum. Nevertheless a two
temperature plasma model with a metal abundance and two plasma
temperatures frozen at the values above ($Z = 0.21$, $kT_1 = 1.02$~keV
and $kT_2= 2.82$) provides a good fit to the \chandra\ data ($P=0.72$)
by letting the two emission measures and the absorbing column density
vary -- see Table~\ref{tab:xmm74ks_ps}. The fitted values for the
two emission measures are $E\!M_1 = (4.3 \pm 1.6) \times 10^{53}~{\rm
cm^{-3}}$ and $E\!M_2 = (2.2 \pm 0.4) \times 10^{53}~{\rm cm^{-3}}$,
consistent with the values derived from the \xmm\ spectrum. The
relative model flux level is $7.3 \times 10^{-13}$\ecms\ (in the
$0.55$--$7.50$ keV band). The value for the absorbing column density
derived in this way from the \chandra\ data, $N({\rm H}) = (1.24\pm
0.09) \times 10^{22}$~cm$^{-2}$ is higher than the value derived from
\xmm\ (and a model with the value derived from \xmm\ does not provide
an acceptable description of the data). We note though that a
systematically higher value for the absorbing column density derived
from \chandra\ data is consistent with the known presence of (likely
carbon-based) contamination on the ACIS chips.  This causes additional
low-energy absorption (up to 50\% near the C edge) not accounted for
in the current response matrices (\citealp{psm+02}).
 
Given the source variability during these $\sim 64$~ks of observation,
we have investigated the presence of significant spectral variation.
We have subdivided the data into three intervals (the first 30 ks, the
following 18 ks and the final 16 ks), chosen to ensure similar
statistics in the resulting spectra. As the initial part of the
observation is rather heavily contaminated by high background, a
higher fraction of it was discarded, and therefore the first interval
is significantly longer. The three spectra were modeled with an
absorbed two temperature plasma model. The fitted values of the model
parameters are summarised in Table~\ref{tab:xmm74ks_ps}. As it can be
seen by inspecting the table, the short term variability observed in
the system V892~Tau during this first \xmm\ observation does not
appear to be associated with significant spectral changes.

\begin{figure}[!tbp]
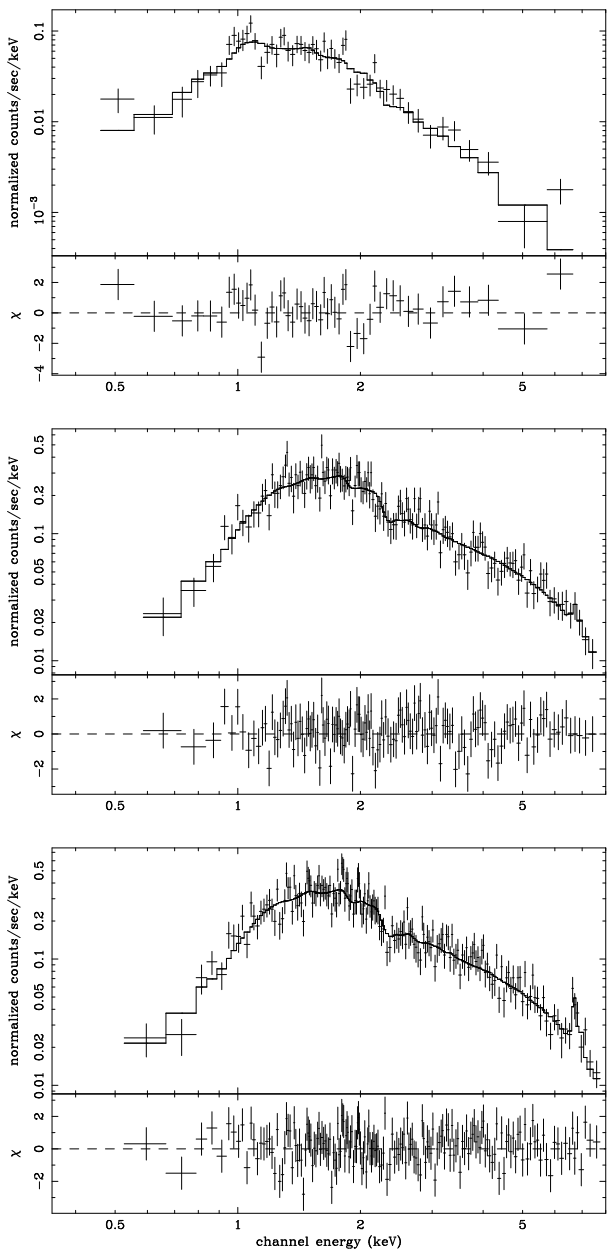

  \begin{center} \leavevmode 
	\epsfig{file=151f8.ps, height=8.0cm, angle=270, bbllx=80, bblly=43, bburx=535, bbury=700, clip=}
	\epsfig{file=151f9.ps, height=8.0cm, angle=270, bbllx=80, bblly=43, bburx=535, bbury=700, clip=}
	\epsfig{file=151f10.ps, height=8.0cm, angle=270, bbllx=80, bblly=43, bburx=554, bbury=700, clip=}

\caption{Background-corrected spectra of the V892~Tau system during
  the tree different phases of the second \xmm\ observation. From top
  to bottom: 14 ks integration time before the flare, 8 ks during the
  rising phase of the flare, 10 ks during the flare maximum.}

  \label{fig:ps_xmm}
  \end{center}
\end{figure}

A time resolved spectral analysis was also performed for the second
\xmm\ observation of V892~Tau system, when the large flare took place.
The data were subdivided into three intervals: a first 14 ks interval
while the source is quiescent, a second 8 ks interval while the source
counts are rising and the last 10 ks, while the source is at its
luminosity maximum\footnote{The segment lengths do not add up to 45
ks, because of the gaps in the data due to the filtering out of the
solar proton flares.}. The three spectra are shown in
Fig.~\ref{fig:ps_xmm}, and their best-fit parameters are listed in
Table~\ref{tab:spectra}.  As explained at the beginning of this
section an absorbed one-temperature plasma model provides an
acceptable description for all the three spectra derived from this
exposure, so we did not attempted fits with two-temperature plasma
models (which were necessary to obtain acceptable fits of the spectra
derived from the first \xmm\ exposure).  In addition, the approach to
flare modelling that we present in Sect.~\ref{sec:flare} relies on
single temperature paramaterization of the X-ray spectrum.
Nevertheless, for the spectrum derived from the first 14 ks interval
of this exposure (corresponding to the quiescent phase), we verified
that an absorbed two temperature plasma model with a metal abundance
and two plasma temperatures frozen at the values derived from the
first \xmm\ exposure ($Z = 0.21$, $kT_1 = 1.02$~keV and $kT_2= 2.82$)
indeed fits the spectrum ($P=0.08$). The value for the absorbing
column density derived in this way is $N({\rm H}) = (0.83\pm 0.08)
\times 10^{22}$~cm$^{-2}$, somewhat higher than the value derived with
the one-temperature plasma fit ($N({\rm H}) = (0.65\pm 0.06) \times
10^{22}$~cm$^{-2}$, second line of Table~\ref{tab:spectra}), and in
better agreement with the value derived from the first \xmm\
exposure. This confirms the lack of significant spectral variation in
the V892~Tau system during the quiescent phase.

As can be seen from Table~\ref{tab:spectra}, during the flare the only
significant spectral variation occurs in the plasma temperature, while
the fitted values for absorbing column density and plasma metallicity
remain essentially unchanged.  The plasma temperature increases from
$kT = 1.53$~keV before the flare to $kT = 8.11$ during the rising
phase and remains around that value afterwards. The model dependent
flux density of the source, in the energy band $0.35$--$7.50$~keV,
goes from $6.7\times 10^{-13}$~\ecms\ before the flare to $1.0\times
10^{-11}$~\ecms\ during the flare maximum. The peak X-ray luminosity
for the flare is $L_{\rm X} = 2.4 \times 10^{31}$\es, high but not
exceptionally so for stellar X-ray flares.


\begin{table*}[thbp]

\caption{The first four lines of the table report the best-fit
spectral parameters for the V892~Tau system during the first \xmm\
observation (before the flare). The parameters are derived assuming a
two-temperature optically thin plasma model. Fits with a single
temperature absorbed plasma model are unacceptable in this case.
The last line reports the result of a spectral fit to the \chandra\ data
on V892~Tau with a two-temperature plasma model with the two plasma
temperatures and the metal abundance frozen at the values derived from
this \xmm\ exposure (first line). See text.}

\begin{center} 	
{\small    \begin{tabular}{ccccccccc}

Time interval & $N({\rm H})$ & $kT_1$ & $kT_2$ &
$E\!M_1$ & $E\!M_2$ & $Z$ & $\chi^2$ & $P$ \\
\hline
ks & $10^{22}~{\rm cm^{-2}}$ & keV & keV & $10^{53}$  cm$^{-3}$
& $10^{53}$  cm$^{-3}$ & Z$_\odot$ & ~ & ~\\
\hline

0--64 (total) & $0.92\pm 0.05$ & $1.02\pm 0.04$ & $2.82\pm 0.41$
& $5.62\pm 2.61$ & $3.56\pm 1.21$ & $0.21\pm 0.06$ & 1.07 & 0.21\\

0--30  & 0.94$\pm$0.10 & 0.98$\pm$0.10 & 2.69$\pm$0.32 & 2.65$\pm$2.16 &
4.46$\pm$1.09 & 0.51$\pm$0.20 & 0.92 & 0.66\\

30--48 & 0.89$\pm$0.07 & 1.08$\pm$0.09 & 3.17$\pm$1.26 & 7.42$\pm$4.98 & 
3.03$\pm$3.08 & 0.13$\pm$0.09 & 1.14 & 0.15\\

48--64 & 0.98$\pm$0.13 & 0.77$\pm$0.08 & 1.95$\pm$0.17 & 3.79$\pm$2.84 &
4.43$\pm$1.11 & 0.31$\pm$0.15 & 1.17 & 0.16\\
\hline
0--18$^*$ & 1.24$\pm$0.09 & 1.02 (froz.) & 2.82 (froz.) & 4.3$\pm$1.6 &
2.2$\pm$0.4 & 0.21 (froz.) & 0.88 & 0.72\\
\hline
    \end{tabular}
}
    \label{tab:xmm74ks_ps}
  \end{center}
	$^*$\chandra\ data (total exposure time)
\end{table*}

\begin{table*}[thbp]
  
   \caption{Best-fit spectral parameters for V892~Tau and V892~Tau~NE
	during the \chandra\ observation and the second \xmm\
        exposure. The spectral parameters are derived assuming a
        single-temperature optically thin plasma model.}  
\begin{center} 
\leavevmode
    \begin{tabular}{lcccccc}

Source 	& $N({\rm H})$	&  $kT$ & $E\!M$ & $Z$ & $\chi^2$ & $P$\\
\hline
~ 	& $10^{22}~{\rm cm^{-2}}$ & keV & $10^{53}$  cm$^{-3}$
    &Z$_\odot$ & ~ & ~\\
\hline
V892~Tau (Chandra) & 0.83$\pm$0.08 & 2.10$\pm$0.19 & 4.31$\pm$0.95 &
    0.06$\pm$0.07 & 1.04 & 0.40\\ 

V892~Tau (XMM 1) & 0.65$\pm$0.06 & 1.53$\pm$0.16 &  5.19$\pm$1.49 &
    0.09$\pm$0.05 & 1.33 & 0.06\\ 
V892~Tau (XMM 2) & 0.97$\pm$0.05 & 8.11$\pm$1.00 &  21.22$\pm$1.01 &
    0.16$\pm$0.09 & 0.90 & 0.79\\
V892~Tau (XMM 3) & 1.00$\pm$0.04 & 6.68$\pm$0.58 &  26.44$\pm$1.11 &
    0.26$\pm$0.08 & 1.01 &0.46\\
\hline
V892~Tau~NE & 1.10$\pm$0.21 & 1.08$\pm$0.22 & 0.78$\pm$0.56 & 
     0.2 (froz.) & 0.40 & 0.98\\
\hline
    \end{tabular}
    \label{tab:spectra}
  \end{center}
	XMM 1: X-ray quiescent,
	XMM 2: rising phase of large flare,
	XMM 3: flare maximum
\end{table*}

\section{Discussion}
\label{sec:disc}

During the 18 ks \chandra\ observation, V892~Tau (which is well
resolved from the apparent companion) presents significant
variability, with its X-ray flux varying by a factor of 2 in less than
1 ks.  This type of variability has never been reported before for a
Herbig Ae star. Undoubtedly, though, it is the large variation of
luminosity of the V892~Tau system (the large flare) observed with
\xmm\ that is most remarkable. The steep impulsive rise of the source
counts by a factor of 10 together with the impulsive rise of the
plasma temperature (from 1.5 to 8.1 keV) are all consistent with the
interpretation of the observed variation in terms of a large coronal
flare.

As already mentioned, V892~Tau and V892~Tau~NE are unresolved by the
\xmm\ PSF, nevertheless the evidence that the origin of the observed
flare is the Herbig Ae star is compelling. First, as summarised in
Table~\ref{tab:coord}, the \xmm\ position, before and during the
flare, of the source associated with system the V892~Tau is between 2.4
and 3.1 arcsec from the radio position of the main star V892~Tau,
while is between 6 and 6.8 arcsec off the position of V892~Tau~NE.
The absolute measurement accuracy of the \xmm\ pointing is 4 arcsec
(\xmm\ Users' Handbook), so the offset of the source in the EPIC-pn
image from the radio coordinates of V892~Tau is well within 1~$\sigma$
of the pointing accuracy. This is fully consistent with V892~Tau being the
dominant source of X-ray emission throughout the \xmm\ observation, as
it is the case during the \chandra\ observation, where V892~Tau is
roughly 10 times more luminous than V892~Tau~NE.

Second, the \xmm\ position of the source associated with the system
V892~Tau before and during the flare is constant within 0.7
arcsec\footnote{The 0.7 arcsec shift of position of the source before
  and during the flare is to the south, while V892~Tau~NE is to the
  north-north east of V892~Tau. Therefore, although not significant
  given the $\sim 1$~arcsec accuracy of \xmm\ relative pointing error,
  the position shift is consistent with V892~Tau becoming even stronger in
  respect with V892~Tau~NE.}, while the angular separation of V892~Tau
and V892~Tau~NE is 4.1 arcsec. At the same time (as shown in the
previous section) the spectral characteristic of the EPIC-pn data on
the V892~Tau system while quiescent are in good agreement with the
characteristics derived from the spectral data on V892~Tau from
\chandra\, but significantly different from the ones derived from the
\chandra\ spectral data on V892~Tau~NE.  These two facts strongly
indicate that the \xmm\ data are, for both the quiescent emission and
the flare, fully dominated by the emission of V892~Tau. Had the flare
been associated with V892~Tau~NE, a shift of the X-ray source position
during the flare would have been observed.

Finally, if one wanted to explain the \xmm\ light curves as mostly
determined by V892~Tau~NE, disregarding the evidence coming from the
pointing and spectral information, one would have to require that the
source which does not appear significantly variable in the \chandra\ 
observation becomes from ten (before the flare) to a hundred times
(during the flare) brighter during \xmm\ observation.  This is
unlikely.

Given the facts above, we conclude that the \xmm\ light curve shown in
Fig.~\ref{fig:lc_xmm} and the EPIC-pn spectra shown in
Fig.~\ref{fig:ps_xmm} and Fig.~\ref{fig:ps_xmm74ks} are determined by
the activity of the Herbig Ae star V892~Tau and minimally influenced
by the presence of V892~Tau~NE. Consequently, below we discuss the
\xmm\ data as originating from V892~Tau and do not further comment on
the unresolved presence of V892~Tau~NE.

\subsection{Quiescent emission}

By ``quiescent'' here we mean the time when V892~Tau is not undergoing
the large flare, even though during this phase we do observe smaller
flare-like events (both in \chandra\ and \xmm\ data).

The \chandra\ spectrum and the \xmm\ spectra of V892~Tau while the
source is quiescent can all be well described by an absorbed two
temperature plasma model with $N({\rm H})\sim 0.9 \times 10^{22}~{\rm
  cm^{-2}}$ $kT_1=1.0$~keV, $kT_2=2.8$~keV and a relative metal
abundance of ${\rm Z = 0.2~Z_{\sun}}$. As discussed above, for
the \chandra\ data a slightly higher value of $N({\rm H})$ is required in
order to fit the spectrum, consistent with the known contamination of
the ACIS chips. Here therefore we only refer to the value for $N({\rm
  H})$ derived from the \xmm\ data.

The X-ray emission from V892~Tau has been studied previously by
\cite{ss94} and \cite{zp94}. These two works use the same ROSAT
observation but derive from it different model parameters.
\cite{ss94} fit the spectral data with an absorbed two temperature
plasma model with $N({\rm H})\simeq 1.3\pm0.6 \times 10^{22}~{\rm
  cm^{-2}}$, $kT_1=0.55$, $kT_2=1.20$ and estimate an X-ray luminosity
$L_{\rm X} = 1.0 \times 10^{31}$~\es\ (in the 0.2--2.4 keV band).
\cite{zp94} model the spectra of V892~Tau with an absorbed 1
temperature plasma with $N({\rm H})=(4.8 \pm 1.1)\times 10^{21}~{\rm
  cm^{-2}}$, $kT=2.3 \pm 0.4$~keV and estimate $L_{\rm X} = (1.9 \pm
0.8) \times 10^{30}$~\es\ (a factor of $\simeq 5$ lower than the
estimate of \citealp{ss94}).

We derive $N({\rm H})\sim 0.9 \times 10^{22}~{\rm cm^{-2}}$,
compatible to the estimate of \cite{ss94}. This value for the
absorbing column density corresponds to a visual
extinction\footnote{using a conversion factor $\nh/A_{V}=1.9\times
  10^{21}$ atoms~cm$^{-2}$~mag$^{-1}$ (see e.g.\ \citealp{cox00}).} of
$A_{V} \sim 4.7$, which is close to the value estimated by \cite{zp94}
from the $B-V$ value given in \cite{hb88} and a factor of $\sim 2$
lower than derived by \cite{elias78} and
\cite{ss94}\footnote{\cite{ss94} estimate $A_{V}$ from optical
  observation. This may explain why their value for the absorbing
  column density is close to our estimate while their estimate for
  $A_{V}$ is a factor of 2 higher.}.  We note that a lower value of
visual extinction is consistent with the star spectral type being A6
rather than B9.

The two temperatures for the two component plasma model that we derive
here are significantly higher than the values given by \cite{ss94};
this is not unexpected given the much softer (and narrower) bandpass
of the ROSAT PSPC instrument used by \cite{ss94}. The temperature
derived from our 1 plasma component fit to the \chandra\ data ($kT =
2.1$~keV) is however consistent with plasma temperature estimated by
\cite{zp94}. The values we derived for the two plasma temperatures in
V892~Tau are somewhat higher than the typical values derived for the
X-ray emission from low mass PMS in Taurus: in an analysis of the
spectral characteristic of 9 T Tauri stars in L1551 we found typically
$kT_1 \sim 0.3$ and $kT_2 \sim 1.2$ (\citealp{fgm+03}).  On the other
hand the value for the plasma metallicity $Z=0.2$ for V892~Tau is
typical of the value we derived in the same study for the 4 Weak Lined
T Tauri stars in the sample.

The intrinsic luminosity we estimate for V892~Tau in its quiescent
state is $L_{\rm X}=1.6\times 10^{30}$~\es, in good agreement with the
value given by \cite{zp94} and a factor of $\sim6$ lower than derived by
\cite{ss94}.  This indicate that \cite{zp94} are probably correct when
they suggest that the reason for the discrepancy between their values
and the one derived by \cite{ss94} is the fact that Strom \& Strom may
have erroneously included the nearby source [BHS98] MHO 11 in the
source circle of V892~Tau.

A wind-related origin of the X-ray emission has been proposed
for Herbig Ae/Be stars by \cite{zp94} and \cite{dms+94}.  This
scenario, however, seems an unlikely explanation for the origin of the
X-ray emission on V892~Tau in its quiescent phase.

As described in Sect.~\ref{sec:analysis}, during this phase, V892~Tau
presents significant short term variability, with its X-ray flux
varying by a factor of 2 in less than 1 ks. The impulsive rise in flux
appears to be associated with an hardening of the spectrum in the
\chandra\ data.  This type of variability is similar to the one
observed in lower-mass pre-main-sequence stars, where the X-ray
emission is of coronal origin, while being substantially different
from the 20--30\% X-ray flux variations observed in OB stars
(\citealp{css+89}; \citealp{ccm+94}) where the emission mechanism is
wind-related.

In addition, from the luminosity of V892~Tau estimated to be around
$38~L_{\sun}$ (\citealp{bci+92}), its ratio of X-ray luminosity to the
total luminosity while quiescent is $L_{\rm X}/L_{\rm bol} \simeq
1\times 10^{-5}$. This is two orders of magnitudes greater than the
typical ratios found for OB stars. On the other hand a value of
$L_{\rm X}/L_{\rm bol} \simeq 10^{-5}$ is not uncommon for low mass
stars in which the X-ray emission mechanism is coronal.

The possibility of a wind related origin of the X-ray emission from
V892~Tau in its quiescent phase appears therefore unlikely.  This
is in agreement with the conclusions reached in the two statistical
studies of the properties of Herbig Ae/Be stars of \cite{pz96} and
\cite{hky+02}. These studies indicate that the X-ray emission from
Herbig Ae/Be stars is generally associated with higher plasma
temperature and higher X-ray to bolometric luminosity ratios than
typically observed for the X-ray emission from main sequence OB
stars.

A wind scenario, finally, cannot account for the observed flare
event of V892~Tau.

\subsection{Flare event}

During the flare event the X-ray luminosity of V892~Tau increases by a
factor of $\sim 15$, from $L_{\rm X}=1.6 \times 10^{30}$~\es\ to
$L_{\rm X}=2.4 \times 10^{31}$~\es, while the temperature of the
plasma increases from $kT=1.5$~keV to $kT=8.1$~keV. The source
luminosity increases over a relatively long time of $\simeq 10$~ks,
and hovers close to the luminosity maximum for at least another 10
ks thereafter -- the end of the observation does not unfortunately
allow to study the decay phase.  As demonstrated by \cite{rbp02}
flares of this type cannot take place in the freely expanding
plasmoids of a stellar wind. The observed slow increase of X-ray
luminosity at the beginning of the flare event requires the presence
of a confining magnetic field.

In order to gain some quantitative insight on the event and given the
similarities of the derived light curve with the ones of other stellar
flares we have analyzed the event through scaling obtained from
detailed hydrodynamic models of flaring plasma confined in a closed
coronal loops, as in solar flares.

\subsubsection{Flaring region characteristics}
\label{sec:flare}

It is customary to derive information about the size of the flaring
loops from the flare evolution, i.e. the light curve. It has been
shown that the decay time of the light curve is linked to the plasma
cooling times, which, in turn, depends on the length of the loop which
confines the plasma (e.g. \citealp{rea2002} and references therein):
the slower the decay, the longer the loop, unless a significant
residual heating sustains the decay and makes the
decay-time/cooling-time dependence less tight.

The \xmm\ observation of the flare on V892 Tau does not cover the
decay phase and therefore diagnostics using the characteristics decay
times are not feasible. For this particular flare, however, we are
able to infer some information on the size of the flaring region from
the observed rise phase.

The evolution of the flaring plasma confined in a loop is well-known
from extensive hydrodynamic loop modeling (e.g. \citealp{psv82}): a
strong heating pulse is triggered in an initially quiescent coronal
loop and makes the temperature increase by up to several tens of MK
along the whole loop in a few seconds, due to the high plasma thermal
conduction. The dense chromosphere at the loop footpoints is heated
violently and expands upwards with a strong evaporation front.  The
rising plasma fills up the loop, very dynamically first and then more
gradually, approaching a new hydrostatic equilibrium at a very high
pressure. The loop X-ray emission increases mostly following the
increase of emission measure, and forms the rise phase of the flare.

Although the evolution in the rise phase is very dynamic and
non-linear, we can nevertheless derive an approximate time scaling.
From the equation of energy conservation of the confined plasma (e.g.
Eq.(3) in \citealp{srj91}) it can be seen that, after the very initial
seconds, dominated by the plasma kinetics, the evolution in the bulk
of the rise phase can be approximately described as a linear increase
of the plasma internal energy density driven by the (constant) energy
input rate per unit volume:

\begin{equation}
  \frac{d {\epsilon}}{d t} 
  \sim E_H
\label{eq:ene}
\end{equation} 

The pressure increases linearly as well, with a rate $p^{\prime}
\simeq 2/3 E_H$. The linear trend goes on until other energetic terms,
and in particular the plasma radiative losses, become important in the
energy balance. Then the pressure increase slows down to approach
asymptotically the equilibrium condition described by the loop scaling
laws (\citealp{rtv78}) for a loop with heating volume rate $E_H$. A
constantly linear pressure increase with a rate $p^{\prime}$ would
imply that the loop reaches the equilibrium pressure in a time $\tau_p
\approx p_0/p^{\prime} = 3/2 p_0/E_H$. 

Detailed hydrodynamic simulations of the flaring loop rising phase have
been performed on purpose for the present work. We simulated the
evolution of the plasma confined in a coronal loop brought to
flare conditions by a strong impulsive heating, as
done in previous extensive modeling of solar and stellar coronal flare
(Peres et al. 1982, 1987; Reale et al. 1988; 
Reale \& Peres 1995; Betta et al. 1997, 2001).
The assumption specific for this work is that of a constant flare
heating lasting long enough to let the loop reach equilibrium conditions.
The simulations
show that the pressure linear
increase rate is $\sim 20$\% slower than $p^{\prime}$. If we also
include the slowing of the pressure increase at later times, we obtain
that the time taken to reach $\sim 90$\% of the equilibrium pressure
is twice $\tau_p$, i.e.  $\delta t_p \approx 3 p_0/E_H$. \cite{srj91}
have shown that $\tau_S = p_0/E_H \sim 120 L_9/\sqrt{T_7}$ is the loop
thermodynamic decay time, where $L_9$ is the loop half-length in units
of $10^9$ cm and $T_7$ is the plasma maximum temperature in units of
$10^7$~K. Therefore, we end up with an expression $\delta t_p \sim 3\,
\tau_S$ which links the time taken by the loop to approach the flare
equilibrium condition to the loop thermodynamic decay time.

We note that in this flare the rise phase is relatively long ($\sim
8$~ks), and the flaring loop may therefore approach equilibrium
condition.  So, if we set $\delta t_p$ equal to the rise duration
time of $\sim 8\times 10^{3}$~s, we obtain an estimate for $L_9$,
once $T_7$ is known. In this case $T_7 \sim 18$\footnote{\cite{bg2003}
give the empirical expression $T_{\rm max}=0.184~T_{\rm
obs}^{1.13}$ linking the temperature obtained from spectral fitting of
EPIC data ($T_{\rm obs}$) to the loop maximum temperature ($T_{\rm
max}$)}, so that we obtain $L_9 \sim 100$.

This estimate for the size of the flaring region has been obtained
under a few assumptions which deserve some comments. First, we have
considered a flare occurring in a single loop.  This may not hold true
for such an intense and long-lasting flare, which may perhaps be
described more properly as a two-ribbon flare, consisting of
progressively reconnecting higher and higher loops. Nevertheless, it
often occurs in solar two-ribbon flares that the rise phase mostly
involves a dominant loop structure, and then gradually extends to
others (e.g.\ \citealp{aa2001}). We may associate the estimated length
to such dominant structure.

Another implicit non-trivial assumption is that of a heating pulse
constantly high during the rise phase. Indeed, a gradually increasing
heating function may drive the observed gradual rise of the light
curve, invalidating the estimations made above. However, in such a
case we should observe also a gradual increase of the temperature,
while the indications are for a sudden jump of the temperature to the
flare value, which is more typical of a heating pulse.  We may be
therefore quite confident that the total loop length is $\simeq 2
\times 10^{11}$ cm.

If we assume a loop aspect $R/L \sim 0.1$, where $R$ is the radius of
the loop cross-section, assumed circular and constant along the loop,
we obtain a total loop volume $V \sim 6 \times 10^{31}$ cm$^3$; a
maximum emission measure of $\sim 2.6 \times 10^{54}$ cm$^{-3}$ then
implies a maximum average loop plasma density of $\sim 2 \times
10^{11}$ cm$^{-3}$ and a maximum pressure of the order of $10^4$ dyne
cm$^{-2}$.  This value is compatible with the equilibrium pressure
obtained from loop scaling laws (\citealp{rtv78}) and consistent with the
hypothesis of the loop at maximum X-ray luminosity being close to
equilibrium conditions.  In order to confine a plasma at such a
pressure, a magnetic field of more than $\sim 500$ Gauss is required.
The origin of such a field in stars that are thought to be fully or
nearly-fully radiative is puzzling.




\subsubsection{What is the origin of the magnetic field?}

In low-mass stars the presence of a significant convection zone
supports the dynamo mechanism that can generate the confining magnetic
fields at the origin of the X-ray flare events, but according to
classical models, pre-main-sequence star with masses in excess of
2~$M_{\sun}$ are expected to follow fully radiative tracks once the
quasi-static contraction has ended.

\cite{ps90} have made the suggestion that the surface activity and
winds observed in Harbig Ae/Be may be related to the presence of an
outer convection zone. In their interpretation, this convection zone
results from the subsurface shell burning of residual deuterium which
was accreted during the protostar phase. Nevertheless \cite{ps93}
reconsider their hypothesis, and explain that according to their
models the retreat of the proto-star outer convection zone does last a
substantial fraction of the pre-main-sequence lifetime of an
intermediate mass star. During this retreat, however, the effective
temperature remains relatively low, so that the star would not be
observed with an A or B spectral type. They conclude that the presence
in Herbig Ae/Be stars of surface activity and strong winds is not
linked to an outer convection zone, since their model shows that such
convection always vanishes with the rising effective temperature.

Recently \cite{sdf00} have presented calculations of pre-main-sequence
evolutionary tracks for low- and intermediate-mass stars.  These
models predict the existence of a thin convective envelope in young AB
stars.  In their review, \cite{fm03} suggest that this thin convective
envelope, of roughly $2 \times 10^{-3}$ times the stellar radius,
could be at the origin of a low-coronal activity (at the level of the
observed minimum for solar type stars) in Altair (A7V) and thus
explain the source X-ray luminosity of $L_{\rm X} = 3 \times
10^{27}$\es.

According to the same models a star with mass 1.9~$M_{\sun}$, the same
metallicity of the Sun and an age of 10 million years, would have a
spectral type A6, a luminosity of 12~$L_{\sun}$, not too far from the
one of V892~Tau, and a convective envelope of a fraction of 0.0018 its
radius. Since the model predicts a stellar radius of $1.2 \times
10^{11}$~cm, the size of the convective region would be $2.2\times
10^8$~cm ($\sim$ 0.3\% $R_{\sun}$).  It is not clear whether such a
thin convective envelope, that may be sufficient to generate the low
coronal activity invoked to explain the 3 to 4 order of magnitude
fainter X-ray emission of a main sequence A star, can sustain the
dynamo action necessary to explain the strong X-ray activity of
V892~Tau. Indeed, from our flare model we derive a flare loop
length of $\simeq 2\times 10^{11}$ cm. This is comparable to the
stellar radius and corresponds to a size of $\sim 500$ times the thin
convective envelope.

An alternative mechanism to sustain the dynamo activity in these
predominantly radiative stars has been proposed by \cite{tp95}. They
argue that dynamo activity can be sustained in AB stars for a
substantial fraction of their pre-main-sequence life time by tapping
the initial stellar differential rotation -- or shear energy.

\section{Conclusions}
\label{sec:concl}

We have analysed the light curves and spectral data of the system
V892~Tau and V892~Tau~NE in a \chandra\ 18 ks exposure and 2 
consecutive \xmm\ exposures of 74 and 45 ks (nominal).  In the \chandra\ 
data, the Herbig Ae star V892~Tau is well resolved from the low mass
later type apparent companion V892~Tau~NE. During the \chandra\ 
observations V892~Tau shows significant variability, with its X-ray
flux varying by a factor of 2 in less than 1 ks. This type of
variability is reminiscent of the type of flaring variability observed
in lower-mass pre-main-sequence star and has never been reported
before for a Herbig Ae star.

During the second \xmm\ exposure of the system V892~Tau and
V892~Tau~NE a large flare event takes place. The source luminosity
impulsively increases by a factor of 15, from $1.6\times
10^{30}$~\es\ to $2.4\times 10^{31}$~\es, while the temperature of the
plasma increases from $kT=1.5$~keV to $kT=8.1$~keV.

V892~Tau and V892~Tau~NE are unresolved by the \xmm\ PSF, nevertheless
the combined \xmm\ and \chandra\ data set provide strong evidence that
the origin of the observed flare is the Herbig Ae star.

We have modeled the flare event and find that a magnetic field of 500
Gauss in intensity is necessary to confine the plasma which reaches
a temperature in excess of 100~MK. The generally accepted mechanism
that can sustain confining magnetic fields of this intensity is a
dynamo action. Therefore under the assumption that surface convection
is a necessary ingredient for a dynamo, our findings imply the
presence of a convective envelope in the Herbig Ae star V892~Tau.

Recent models by \cite{sdf00} indicate that a pre-main-sequence A6
star similar to V892~Tau has a very thin convection zone, extending to
a depth of only 0.2\% of the stellar radius. Whether this thin
convective envelope is sufficient to sustain the dynamo action
required to explain the vigorous X-ray activity of this Herbig Ae star
remains to be assessed.

\begin{acknowledgements}
  
  This paper is based on observations obtained with \xmm, an ESA
  science mission with instruments and contributions directly funded
  by ESA Member States and the USA (NASA). GM and FR acknowledges
  support by Agenzia Spaziale Italiana and by Ministero della
  Universit\`a e della Ricerca Scientifica e Tecnologica.

\end{acknowledgements}

\bibliographystyle{aa}

\end{document}